\documentclass{PoS}     
\newcommand{\ts}{\thinspace}

\newcommand{\E}{1E{\ts}1740.7$-$2942}
\newcommand{\gtsim}{\lower.5ex\hbox{$\; \buildrel > \over \sim \;$}}
\newcommand{\ltsim}{\lower.5ex\hbox{$\; \buildrel < \over \sim \;$}}
\title{
       Comptonization in 1E{\thinspace}1740.7$-$2942 spectra \\
       from 2 to 200{\thinspace}keV
      }

\ShortTitle{Comptonization in 1E{\thinspace}1740.7$-$2942}

\author{
        \speaker{Manuel Castro$^a$}, Ta\'is Maiolino$^a$, Flavio D'Amico$^a$, J\"orn Wilms$^b$,\phantom{@@@@@@@@@}
	Katja Pottschmidt$^c$ and Jo\~ao Braga$^a$ \\
      \llap{$^a$} INPE, Av. dos Astronautas 1758, 12227-010 S.J.Campos-SP, Brazil\\
      \llap{$^b$} Remeis Observatory \& ECAP, Univeristy of Erlangen-Neremberg, Germany\\
      \llap{$^c$} CRESST, NASA/GSFC \& UMBC, USA                 \\
      E-mail:\email{<castro|tais.maiolino|damico|braga>@das.inpe.br}
	    \email{joern.wilms@sternwarte.uni-erlangen.de}
	    \email{katja@milkyway.gsfc.nasa.gov}
       }

\abstract
         {
          Studies of the long-term spectral variations have been used to
          constrain the emission processes of black hole candidates. However, a
          common scenario which is able to explain the emission from soft to
          hard X-rays has been proposed only recently.  Here, we use XMM and
          INTEGRAL data on \E\ in order to demonstrate that Comptonization plays
          an important role in producing high energy photons, as predicted by
          the current modeling scenario.  
         }

\FullConference
               {
                An INTEGRAL view of the high-energy sky (the first 10 years)"
                9th INTEGRAL Workshop and celebration of the 10th anniversary of the launch,\\
		October 15-19, 2012                                                         \\
		Bibliotheque Nationale de France, Paris, France
               }

\begin{document}

\section{Introduction}

\E\ is a black hole candidate (BHC) discovered by the
\textit{Einstein} satellite \cite{Hertz1984}, and it is one of the
brightest hard X-ray sources close to the centre of our Galaxy
\cite{Belanger2006}.  The source has a very hard X-ray spectrum and is
known to be associated with a doubled-sided radio-emitting jet
\cite{Mirabel1992}. For this reason, \E\ is classified as a
\textit{microquasar}.  Due to the high Galactic extinction towards the
Galactic Centre, so far it has not been possible to identify a stellar
companion, but some observations suggest that the companion might be a
low-mass star \cite{Marti2000}. Many studies have been carried out to
understand the nature of the source and the origin of its emission.
In this work, we present a spectral analysis of \E\ using data from
the XMM and INTEGRAL satellites in two epochs: 2003 and 2005. The data
from XMM, to our knowledge, has never been reported in the literature.

\section{Data selection, analysis and results}

Data reported here are a first step of an ongoing project which aims
to study long-term spectral variability of {\E\ using INTEGRAL
  \cite{1} and XMM \cite{2}. Here we selected two simultaneous
  observations of {\E} with XMM and INTEGRAL, one in 2003 and another
  in 2005.  Data were selected from the PN \cite{pn} camera of XMM and
  IBIS/ISGRI \cite{4} telescope onboard INTEGRAL.  INTEGRAL spectral
  extraction was done with the use of the OSA{\ts}9.0 \cite{5}
  software. XMM spectral extraction was done with the SAS software
  version 1.52.9 \cite{sas}.  Data from PN camera were then analyzed
  in the 2--10{\ts}keV band. Following the data analysis threads from
  SAS, we checked if the data were affected by pile-up, concluding that
  no corrections were necessary in the data.  We want to mention that
  XMM data below 2{\thinspace}keV were not included in this
  preliminary work for two main reasons: the residuals of our fits
  (when the part below 2{\thinspace}keV is included) show a presence
  of a feature that can be modeled with a Gaussian line. It is still
  not clear to us if this is due to an instrumental effect or
  not. Including the data from below 2{\thinspace}keV would allow us
  to better constrain the {\tt nH} parameter, but being not sure of the nature
 of the low-energy feature, and adopting an {\tt nH} value from previous measurements,
we proceeded with a lower energy threshold
  of 2{\thinspace}keV.

IBIS data lower threshold starts at 20{\ts}keV extending up to
200{\ts}keV. Data were analyzed with the use of XSPEC.  Such coverage
allowed us to test the framework currently used to model the spectra
of black hole candidates \cite{6}. It is noteworthy that in this first
step we decided not to use JEM-X data to fill the gap between XMM and
INTEGRAL data.  Data were fitted, as suggested by Bouchet et al. 
\cite{6}, with a single (absorbed) Comptonization model. In XSPEC
(language) we used the {\tt comptt} model \cite{8}. It is interesting
to note that more specialized models such as {\tt compps} \cite{9} are
also adequate to fit the spectra. A normalization factor between XMM
and INTEGRAL was also included in the fit, and left free to vary in
the 2003 data. The same normalization factor (equal to 0.9) was used
(and frozen) in the 2005 data. As already mentioned, starting at
$\sim${\ts}2{\ts}keV, XMM data were not well suitable to estimate the
hydrogen column ({\tt nH} parameter in XSPEC). This value was frozen
at $10^{23}$\,cm$^{-2}$ \cite{Gallo}.  Our
spectral fits are shown in Fig. \ref{fig1}. 
We used the disk geometry and optical depth frozen at a value of 1
  \cite{6}. The values of the fluxes and fit parameters 
are shown in Tab. \ref{fit1}.

\begin{table}[h!]
\centering
\begin{tabular}{cccccc}
\hline 
\hline
Period & $\chi^2_{red}$ & kT$_0$ & kT$_e$ & Flux       & Flux\\ 
       &  (keV) & (keV) &   (2-10\,keV)  & (20-200\,keV)  &\\ 
\hline
\vspace{3mm} 
$2003$ &  1.2 & $1.02^{+0.03}_{-0.03}$ & $63.2^{+2.9}_{-2.8}$ & $2.86^{+0.08}_{-0.06}$ & $1.65^{+0.05}_{-0.03}$ \\  
$2005$ &  1.7 & $0.90^{+0.01}_{-0.01}$ & $53.8^{+1.5}_{-1.5}$ & $2.31^{+0.05}_{-0.05}$ & $1.03^{+0.02}_{-0.02}$\\ 
\hline
\end{tabular} 
\caption{
         The fit parameters kT$_e$ (electron temperature), kT$_0$ (seed Wien photon temperature) 
         and fluxes for the combined PN/XMM and ISGRI/INTEGRAL \E\
         spectra with a (absorbed) Comptonization model ({\tt const*phabs*compTT} in XSPEC). 
         Fluxes are in  units of 10$^{-10}$\,erg\,cm$^{-2}$\,s$^{-1}$ (2-10\,keV)
         and 10$^{-9}$\,erg\,cm$^{-2}$\,s$^{-1}$ (20-200\,keV).
        }
\label{fit1}
\end{table}


\begin{figure}[h!]
\centering
 \includegraphics[width=.8\textwidth]{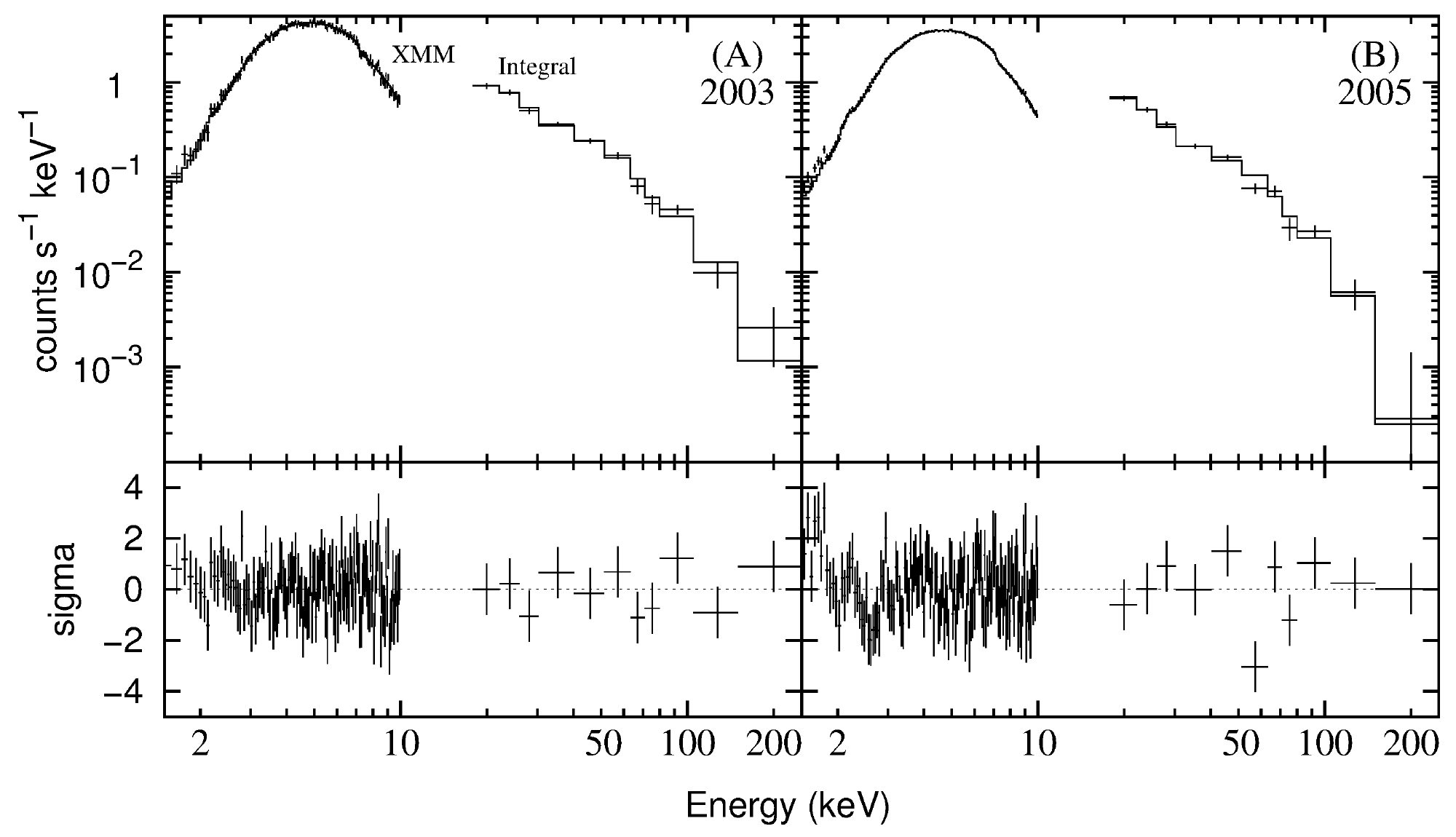}
\caption{
        (A) Data for 2003 and (B) 2005 for \E\ . 
        }

\label{fig1}
\end{figure}


\section{Discussion}

It should be stressed that other model combinations can be used in
order to achieve an adequate fit to the data presented here. As can be
seen from Fig. \ref{fig1} the residuals in the IBIS part can be
interpreted as noisy. Indeed, this is much more clear in the 2005
data, where the residuals of the fit shows that additional components
can be included in the fit. Notwithstanding, the simple
picture shown here with only one component, based on the
Comptonization of lower energy (Wien) photons by a hot plasma (the
scenario of {\tt comptt}) can provide an adequate fit to the data and
is within the framework of modeling the spectra of black hole
candidates \cite{6}. We emphasize that we are claiming that the
fit with only {\tt comptt}, adequate for 2003 data and marginally also for
 2005 data, allowed us to
proceed with (one of) the simplest data modeling, within the reaches and
goals of this study. 
It is tempting to correlate the observed fluxes decrease in both bands
(2-10 and 20-200\,keV) with the variation of kT$_e$. Taking into
account a distance of 7.7$\pm$ 0.4\,kpc to the Galactic Centre
\cite{9}, the observed 2-200\,keV luminosities are, respectively,
$1.4\times10^{37}$ and $9.1\times10^{36}$\,erg/s for 2003 and 2005. It
is not a surprising fact that spectral variations were detected, since
this is a common feature in the spectra of black hole candidates.

\newpage

According to a recent definition of the emission states in black holes
\cite{Remillard}, fits using two components, a powerlaw plus a thermal
(blackbody), are suitable in characterizing such states.  Accordingly,
we also performed fits to our data following this recipe, i.e., using
cutoff powerlaw plus a multicolor disk blackbody model applied to both
spectra (2003 and 2005: see Tab. \ref{fit}). For 2003, a single
powerlaw model (absorbed) was enough to achieve an adequate fit
whereas in 2005 it was necessary to add to the powerlaw a thermal
component ({\tt diskbb}), contributing with less than 0.2{\%} to the
total 2--10{\thinspace}keV flux.  Such a thermal component is
interpreted as originating from the disk and it was already reported in
the spectra of other black hole candidates, such as
GRS{\thinspace}$1758-258$ \cite{14}, as well as in \E\ \cite{15}
\cite{16}.  We were able to determine that in both epochs \E\ was in
the canonical low/hard state.

\begin{table}[h!]
\begin{center}
\begin{tabular}{lrr}
\hline
\hline
 &2003&2005\\ \hline 
$T_{in}$ (keV)&-&0.245$^{+0.040}_{-0.009}$ \\ 
$A_{disk}$&-&48820$^{+50000}_{-39316}$ \\ 
$\Gamma$ &1.38$^{+0.06}_{-0.06}$&1.61$^{+0.03}_{-0.02}$\\ 
$E_{cutoff}$ (keV)&70$^{+18}_{-13}$&76$^{+9}_{-15}$ \\ 
$A_{cut}$&8$^{+1}_{-1}$ $\times10^{-2}$&105$^{+6}_{-6}$ $\times10^{-3}$\\ 
$\chi^{2}_{red}$&0.9&1.2\\ \hline 
\hline
\end{tabular}
\caption{Fit using a power law cutoff model for both epochs. For 2005 the fit also includes a {\tt diskbb} component.}
\label{fit}
\end{center}

 \end{table}

\section{Conclusions}

We present here the first results of an ongoing investigation to study
the long-term variations of \E\ .  Data shown here were collected
during two observations of \E, one carried out in 2003 and the other
in 2005, simultaneously by the XMM and INTEGRAL satellites. This has
provided us a chance to fit the spectrum from 2 to 200\,keV. To our
knowledge these XMM observations are being reported here for the first
time. We fit the spectra with a simple Comptonization.  We observed a
flux variation which correlates with the behavior of kT$_e$ . Our
results are compatible with the current framework of modeling spectra
of black holes candidates with Comptonization from a $\tau\sim1$ and
kT$_{e}\sim60$\,keV plasma, which can explain the origin of the
observed high energy flux (E>20\,keV) as originating from
Comptonization of the lower energy photons.


\begin{thebibliography}{99}

\bibitem{Hertz1984} Hertz, P., Grindlay, J., \textit{The Einstein galactic plane survey - Statistical analysis of the complete X-ray sample},
	\textit{ApJ}, \textbf{278}, 137 (1984)
\bibitem{Belanger2006} Bélanger, G., Goldwurm, A., Renaud, M., et al., \textit{A persistent high-energy flux from the heart of the milky way: INTEGRAL's view of the Galactic Center},
	\textit{ApJ}, \textbf{636}, 275 (2006).
\bibitem{Mirabel1992} Mirabel, F., Rodriguez, L. Cordier, B., et al., \textit{A double-sided radio jet from the compact Galactic Centre annihilator 1E\,1740.7-2942},
      \textit{Nature}, \textbf{358}, 215 (1992)
\bibitem{Marti2000} Mart\'i, J., Mirabel, F., Chaty, S., et al., \textit{VLT search for the infrared counterpart of 1E\,1740.7-2942},  \textit{A\&A}, 
 \textbf{363}, 184 (2000)
\bibitem{1} Winkler, G., Courvoisier, T.J.-L., Di Cocco, G,. et al., \textit{The INTEGRAL mission}, 
	\textit{A\&A},\textbf{411}, L1 (2003)
	
\bibitem{2} Jansen, F., Lumb, D., Altieri, B., et al., \textit{XMM-Newton observatory. I. The spacecraft and operations}, 
	\textit{A\&A}, \textbf{365}, L1 (2001)


\bibitem{pn} Str{\"u}der, L., Briel, U., Dennerl, K., et al., \textit{The European Photon Imaging Camera on XMM-Newton: The pn-CCD camera},
	  \textit{A\&A}, \textbf{365}, L78 (2001)
	
	
\bibitem{4} Ubertini, P., Lebrun, F., Di Cocco, G., et al., \textit{IBIS: The Imager on-board INTEGRAL}, 
	\textit{A\&A}, \textbf{411}, L131 (2003) 
	
\bibitem{5} Goldwurm, A., David, P., Foschini, L., et al., \textit{The INTEGRAL/IBIS scientific data
analysis}, 
	\textit{A\&A}, \textbf{411}, L223 (2003)
\bibitem{sas} Gabriel, C., Hoar, J., Ibarrra, A., et al., \textit{The XMM-Newton SAS - Distributed Development and Maintenance of a Large Science Analysis System: A Critical Analysis},
	\textit{ASP Conference Series}, \textbf{314}, 759 (2004)

	
\bibitem{6} Bouchet, L., Del Santo, M., Jourdain, E., et al., \textit{Unveiling the high energy tail of
1E 1740.7-2942 with INTEGRAL}, 
	\textit{ApJ}, \textbf{693}, 1871 (2009)
	
	
\bibitem{8} Titarchuk, L., \textit{Generalized Comptonization models and application to the recent
high-energy observations}, 
	\textit{ApJ}, \textbf{434}, 570 (1994)
	
\bibitem{9} Meyer, L., Ghez, A.M., Sch{\"o}del, et al., \textit{The Shortest-Known-Period Star Orbiting
Our Galaxy's Supermassive Black Hole}, 
	\textit{Science}, \textbf{338}, 84 (2012)

\bibitem{Gallo} Gallo, E., Fender, R., \textit{Chandra imaging spectroscopy of \E}, \textit{Mon. Not. R. Astron. Soc.}, \textbf{337}, 869 (2002)

\bibitem{Remillard} Remillard, R., McClintock, J., \textit{X-Ray Properties of Black-Hole Binaries}, \textit{ARAA}, \textbf{44}, 49 (2006)

\bibitem{14} Goldwurm, A., Israel, D., Goldoni, P., et al.,  \textit{XMM-Newton observation of the black hole 
microquasar GRS 1758-258}, \textit{AIPC}, \textbf{587}, 61 (2001)

\bibitem{15} Smith, D. M., Heindl, W. A., Swank, J. H.,  \textit{Two different long-term behaviors in black 
hole candidates: evidence for two accretion flows?}, \textit{ApJ}, \textbf{569}, 362 (2002)

\bibitem{16} del Santo M., Bazzano, A.,Zdziarski, A. A., et al., \textit{1E 1740.7-2942: temporal and spectral evolution from INTEGRAL and RXTE observations}, \textit{A{\&}A}, \textbf{433}, 613 (2005) 


\end{thebibliography}
\end{document}